\documentclass[twocolumn,english,aps,prl,reprint,superscriptaddress,floatfix]{revtex4-2}
\usepackage[LGR,T1]{fontenc}
\usepackage[utf8]{inputenc}
\usepackage{verbatim}
\usepackage{amstext}
\usepackage{amssymb}
\usepackage{graphicx}
\usepackage{xcolor}
\usepackage{amsmath}
\usepackage[colorlinks=true, urlcolor=blue, linkcolor=blue, citecolor=blue]{hyperref}
\usepackage{xr-hyper}
\makeatletter

\usepackage[english]{babel}

\addto\captionsenglish{%
}

\newcommand{\MBT}{\mathrm{Mn{Bi}_{2}{Te}_{4}}}

\definecolor{lb}{rgb}{0.000, 0.500, 1.000}

\definecolor{ly}{rgb}{1.000, 0.600, 0.000}

\usepackage[none]{hyphenat}

\makeatother

\begin{document}

\title{N\'eel vector switching by magnetization-dependent layer-resolved spin-orbit torque}

\author{Sheng Pan}
\affiliation{Institute for Quantum Science, School of Physical Science and Technology, Soochow University, Suzhou
215006, China}

\author{Weiwei Wang}
\affiliation{Institutes of Physical Science and
Information Technology, Anhui University, Hefei 230601, China.}

\author{Jie-Xiang Yu}
\email{jxyu@suda.edu.cn}
\affiliation{Institute for Quantum Science, School of Physical Science and Technology, Soochow University, Suzhou
215006, China}

\begin{abstract}
Layer-resolved spin-orbit torques offer a route to controlling antiferromagnetic order through local spin-charge conversion. 
First-principles calculations show that the layer-resolved spin Hall conductivity of the antiferromagnetic topological insulator $\MBT$ consists of a magnetization-independent spin Hall term plus two spin anomalous Hall terms that are even in the N\'eel vector yet evolve with the local magnetic order, 
while local inversion asymmetry differentiates surface and inner layers. 
This response enables reversible, field-free switching of the N\'eel vector in finite-septuple-layer films. 
Although the magnetization-independent spin Hall term alone can reverse the order, 
the spin anomalous Hall terms provide magnetization-dependent feedback that redistributes the layer-resolved torques and reshapes the switching trajectory, 
offering a route to tailoring current-driven switching through layer and interface engineering.
\end{abstract}

\maketitle

Manipulation of the magnetization by spin-orbit torque (SOT) stands as a crucial advancement in spintronics due to its merits in realization of ultrafast switching as well as low-power writing operation\cite{nmat2613,Miron2011,Liu2012TaSHE,Reimers2023}.
In general, SOT originates from the spin current (or spin-polarized current) applied to the magnet, and is given by
\begin{eqnarray}
\boldsymbol\tau_{\mathrm{SOT}}&=& - \frac{1}{1+\beta^2}[\mathbf{m}\times(\mathbf{m}\times \mathbf{u}_{s}) +\beta \mathbf{m}\times\mathbf{u}_{s}]
\label{eq:sot}
\end{eqnarray}
where $\mathbf{m}$ is the unit vector for the local moment and $\beta$ is the secondary spin-torque coefficient. 
The spin current vector $\mathbf{u}_{s}=j_s \mathbf{m}_{p}$ is composed by the product of the magnitude of the spin current and the spin polarization direction. 
It is intriguing that the spin current direction is not manifestly present in Eq.~(\ref{eq:sot}).  
In conventional SOT devices based on heavy-metal/ferromagnet (HM/FM) heterostructures, 
the spin current is generated by the spin Hall effect (SHE) in the HM layer\cite{PhysRevLett.109.096602,PhysRevLett.122.077201,ando2008electric,miao2013inverse},
so that the direction is manifested in the spin-injected direction to the FM layer, as well as in its decay with thickness.
In magnets that spontaneously generate SOT\cite{li2025magnetization,Guo2025,skinner2015complementary}, the direction of the spin current produced by the intrinsic SHE becomes irrelevant.
Instead, the magnitude and polarization direction of the spin current determine the resulting SOT of which the spin current vector $\mathbf{u}_{s}$ given by

\begin{eqnarray}
\mathbf{u}_{s}&=& \boldsymbol\sigma_{xy}^{\mathrm{SHE}} \mathrm{sgn}(\mathrm{\sigma}_{xy}^{\mathrm{AHE}}) \mathcal{E}_{y}
\label{eq:spn-current}
\end{eqnarray}

where $\boldsymbol\sigma_{xy}^{\mathrm{SHE}}$ is the spin Hall conductivity (SHC), 
$\mathrm{\sigma}_{xy}^{\mathrm{AHE}}$ is the anomalous Hall conductivity (AHC), 
and $\mathcal{E}_{y}$ is the applied electric field along the longitudinal direction $y$.
In particular, the characterization of the spin polarization axis of SHE in inherently inhomogeneous magnets such as antiferromagnets (AFM) is both crucial and challenging\cite{PhysRevLett.113.157201,PhysRevLett.119.187204}.
The low conductivities and ultrafast spin dynamics of many AFM 
render them not only outstanding SOT sources, but also ideal tunable media\cite{Wadley2016ElectricalSwitching,sciadv.aar3566,behovits2023terahertz,PhysRevLett.132.136701,Guo2025}.
However, since the net AHC in most collinear AFM (excluding altermagnet) is canceled to zero, 
the studies on AHC and SHC must be spatial-resolved for the identification of SOT\cite{PhysRevB.104.224414,PhysRevLett.130.216702,fkyr-z5b8}.

\begin{figure}
    \includegraphics[width=1.0\linewidth]{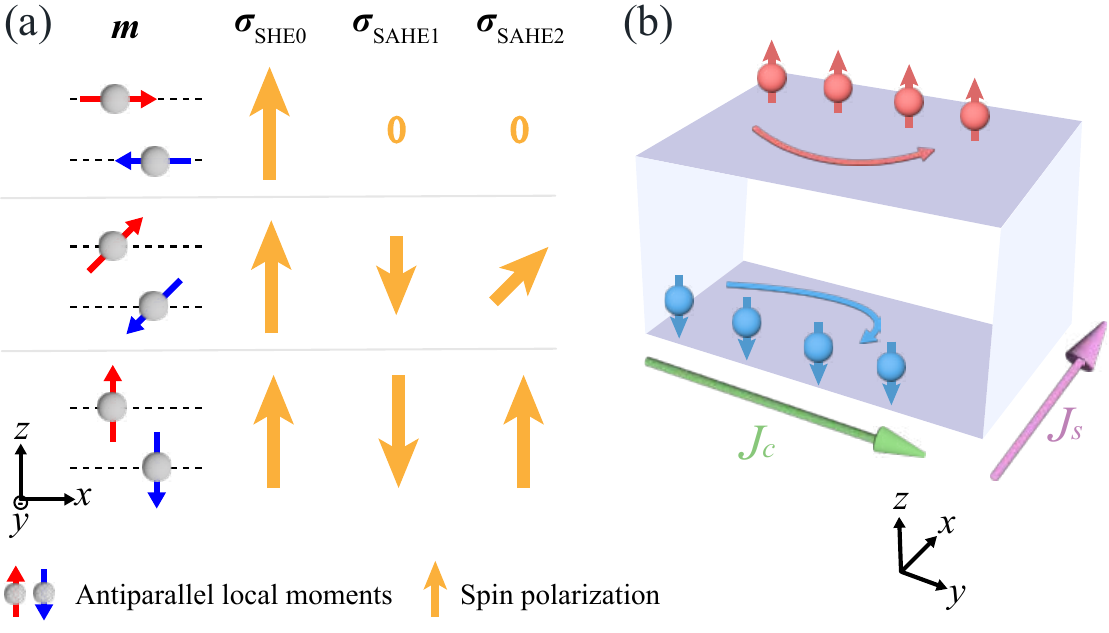}
\caption{
    Schematic of the N\'eel vector orientation-dependent spin response in $\MBT$.
    (a) Response spin polarization in Eq.~(\ref{eq:sahe-sot}). 
    $\boldsymbol{\sigma}_{\text{SHE0}}$ is orientation independent; $\boldsymbol{\sigma}_{\text{SAHE1}}$ changes in magnitude and vanishes for in-plane order, 
    whereas $\boldsymbol{\sigma}_{\text{SAHE2}}$ changes in both magnitude and direction. 
    Red and blue arrows denote antiparallel local moments, and orange arrows denote response vectors. 
    (b) An electric current $\mathbf{J}_{c}$ along $y$ generates a transverse spin current $\mathbf{J}_{s}$ along $x$, with opposite $z$-polarized carriers shown in red and blue.}
    \label{fig:sahe-schematic}
\end{figure}

In this Letter, we investigated the layer-resolved SHE in the AFM topological insulator $\MBT$\cite{otrokov2019prediction,li2019intrinsic,liu2024gatetunable}, 
in which the topological surface states can provide extremely high SOT with small electric current due to the bulk insulating inside the topological band gap\cite{li2025magnetization}.
We discovered the magnetization-dependent SHE where the layer-resolved SHC(LSHC) is given by
\begin{eqnarray}
\boldsymbol{\sigma}_{xy}^{\text{SHE}} &=& \boldsymbol{\sigma}_{\text{SHE0}} \times \hat{\mathbf{x}} \nonumber \\
& &- (\hat{\mathbf{m}} \cdot \hat{\mathbf{z}})^2 (\boldsymbol\sigma_\text{SAHE1} \times \hat{\mathbf{x}}) \nonumber \\
& &+ (\hat{\mathbf{m}} \cdot \hat{\mathbf{z}}) \boldsymbol{\sigma}_{\text{SAHE2}} \odot \hat{\mathbf{m}}
\label{eq:sahe-sot}
\end{eqnarray}
where $\hat{\mathbf{m}}=\hat{\mathbf{m}}_A = -\hat{\mathbf{m}}_B$ is the normalized local moment of AFM with sublattice $A$ and $B$, operator $\odot$ represents Hadamard product.
As illustrated in Fig.~\ref{fig:sahe-schematic}, the first term $\boldsymbol{\sigma}_{\text{SHE0}} \times \hat{\mathbf{x}}$ in Eq.~(\ref{eq:sahe-sot}) represents the conventional SHE contribution, completely independent of the local magnetic state\cite{varotto2020independence,PhysRevLett.131.076702}.
The second and the third term, corresponding to the spin anomalous Hall effect (SAHE)\cite{PhysRevB.99.220405,Iihama2018,Li2022,PhysRevB.99.014403}, are both even functions of $\hat{\mathbf{m}}$, 
indicating both AFM sublattices with opposite sign of magnetization yield the same SHC.
By quantifying the coefficient vectors in Eq.~(\ref{eq:sahe-sot}) as well as the AHC for several-layered systems, 
we then elucidated the distinct behaviors of LSHE-induced SOT (LSOT) in the surface and interior layers of $\MBT$, 
and further demonstrated its field-free robust control via LSOT over the AFM order through spin dynamics simulations.


The electronic band structures of the finite septuple layers (SL) $\MBT$ were obtained by employing density-functional theory (DFT) calculations with projector-augmented wave pseudopotentials implemented in the Vienna $ab$ $initio$ simulation (VASP) package\cite{Kresse_1996_CMS,Kresse_1996_PRB}. 
The generalized gradient approximation (GGA) in Perdew, Burke, and Ernzerhof (PBE) formation is used as the exchange-correlation energy\cite{PAW_1994,PAW_1999}, 
and the Hubbard U method ($U=4.0$ eV, $J=0.9$ eV) is applied on Mn (3$d$) orbitals to include strong-correlation effects\cite{PhysRevLett.122.206401}.
An energy cutoff of 600 eV was used for the plane-wave expansion throughout the calculations. 
The $\Gamma$-centered $9\times9\times1$ $k$-points meshes were adopted for the calculations of few-layer hexagonal lattices with a vacuum space of $ \sim15\,$\AA{}.
For structural relaxations we both relaxed the atoms until the Hellmann-Feynman forces are less than $ 1\, \textrm{meV}/$\AA.
The DFT-D3 method is employed to treat the van der Waals interaction\cite{10.1063.1.3382344}. 
The tight-binding Hamiltonian in a Wannier-function (WF) basis is constructed by performing a unitary transformation implemented in WANNIER90\cite{mostofi2014updated}.
The Kubo formula implemented in Wannier90 code is used to calculate the
k-resolved spin Berry curvature, which is given as\cite{PhysRevLett.49.405, PhysRevB.74.195118}
\begin{equation}
\Omega_{xy,L}^{\gamma}(\mathbf{k})=-\hbar^2\sum_{m\neq n}f_{n\mathbf{k}}(1-f_{m\mathbf{k}})\frac{2{\rm Im}[\hat{j}_{xL,nm}^{\gamma}(\mathbf{k})\hat{v}_{y,mn}(\mathbf{k})]}{(\epsilon_{n\mathbf{k}}-\epsilon_{m\mathbf{k}})^{2}+\eta^2}
\end{equation}
The intrinsic LSHC is calculated using following formulas\cite{qiao2018calculation}
\begin{equation}
\sigma_{xy}^{\gamma}(L)=-\frac{e^2}{\hbar}\frac{1}{VN_{k}}\sum_{\mathbf{k}}\Omega_{xy,L}^{\gamma}(\mathbf{k})
\label{eq:SHC}
\end{equation}
where $f_{n\mathbf{k}}$ represents the Fermi-Dirac distribution function, $\epsilon_{n\mathbf{k}}$, $\eta$ is the smearing parameter and $\epsilon_{m\mathbf{k}}$ are the band energy for each $\mathbf{k}$,
$V$ is the primitive cell volume, and $N_{\mathbf{k}}$ is the number of $\mathbf{k}$ points in the BZ.
For a layer or a group of layers \(L\), the projection operator is defined as
$\hat p_L=\sum_{w\in L}|w\rangle\langle w|$,
where $|w\rangle$ denotes a localized Wannier orbital assigned to $L$ \cite{cyqr-7sgt}. The corresponding layer-resolved Hermitian spin-current operator is
$
\hat j_{xL}^{\gamma}
=
\frac{1}{2}
\left(
\hat p_L\hat s_\gamma\hat v_x
+
\hat v_x\hat s_\gamma\hat p_L
\right)$. Further computational details are provided in the Supplemental Material.

Bulk $\MBT$ is a van der Waals AFM with a rhombohedral structure (space group $R\overline{3}m$) and spatial inversion symmetry.
Each SL follows the Te--Bi--Te--Mn--Te--Bi--Te stacking sequence.
The Mn moments order ferromagnetically within each SL and couple antiferromagnetically between adjacent SLs along the $c$ axis\cite{Otrokov_2017,Gong_2019}.
This out-of-plane $A$-type AFM order breaks the mirror symmetry $M_x$ and time-reversal symmetry $\mathcal{T}$ while preserving the combined symmetry $S=\mathcal{T}\tau_{1/2}$, where $\tau_{1/2}$ is a half-lattice translation.
Each SL also retains a threefold rotation $C_{3z}$ and a mirror plane $\mathcal{M}_{yz}$.

\begin{figure}
\includegraphics[width=1.0\linewidth]{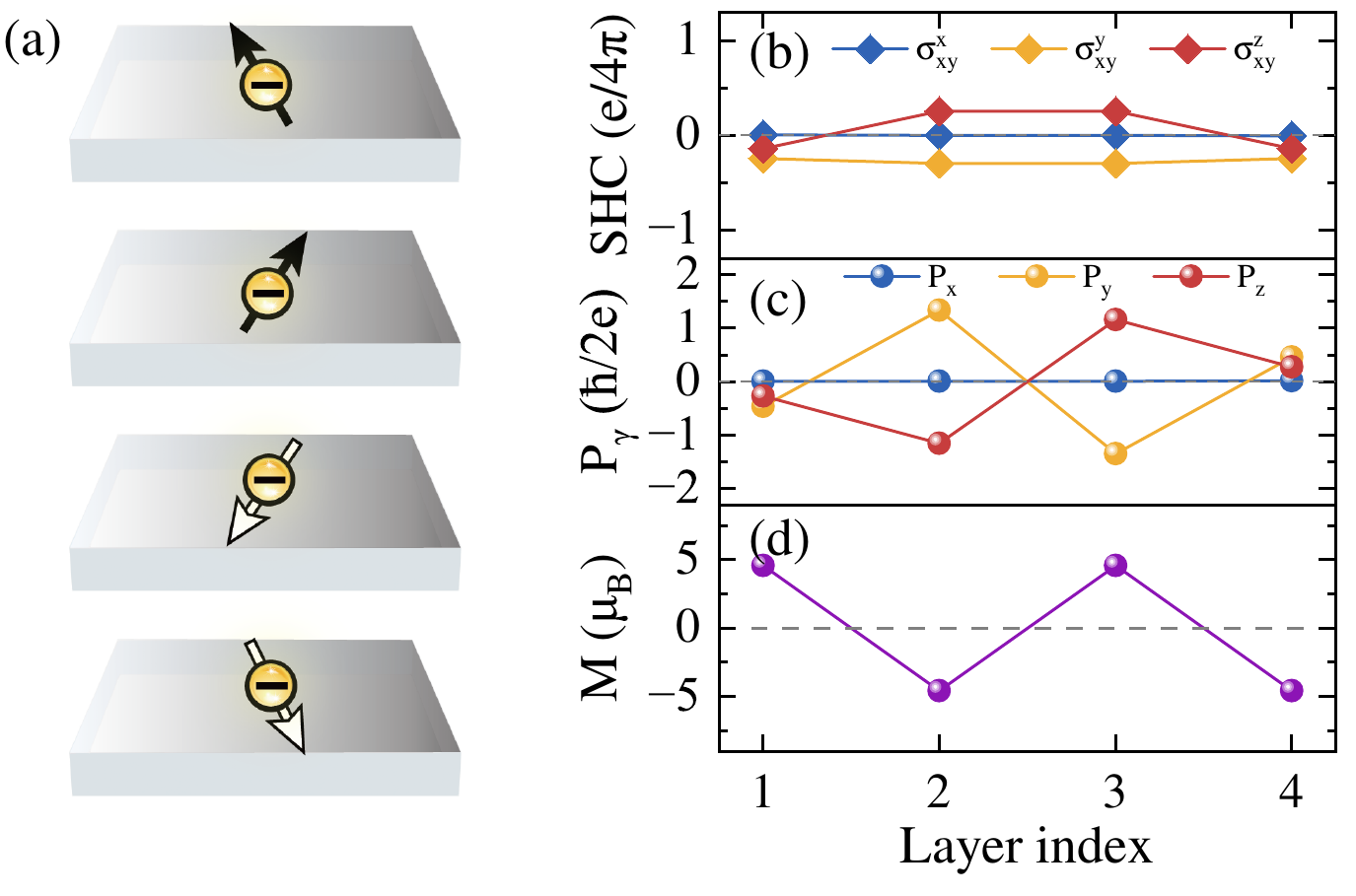}
\caption{In 4 SLs $\MBT$ with the N\'eel vector oriented along the $z$ axis,
(a) schematic of the layer-resolved spin polarization in 4 SLs $\MBT$.
Yellow spheres with arrows represent the local spin polarization directions.
(b) LSHC components $\sigma_{xy}^{x}$, $\sigma_{xy}^{y}$, and $\sigma_{xy}^{z}$ evaluated within the band gap, 
(c) the corresponding layer-resolved spin polarization components, and
(d) layer-resolved magnetic moment $M$.}
    \label{fig:4sls-layer-p}
\end{figure}

For the geometry considered here, an electric field applied along the
$y$ direction generates transverse Hall currents along $x$, as shown in Fig~\ref{fig:sahe-schematic}(b).
To characterize the spin polarization carried by these currents,
we obtained the spin polarization based on layer-resolved AHC (LAHC) and LSHC for each SL via $P(L)=\frac{j_\uparrow(L) - j_\downarrow(L)}{j_\uparrow(L) + j_\downarrow (L)} \approx \frac{\sigma_{xy}^{\gamma}(L)}{\sigma_{xy}(L)}$\cite{rycerz2007valley, cheng2016valley}, where $j_{\uparrow/\downarrow}(L)$ is the layer-resolved spin-polarized current in the spin up/down channels, and $\sigma_{xy}^{\gamma}(L)$ and
$\sigma_{xy}(L)$ are the LSHC and LAHC of layer $L$, respectively. 
Figs.~\ref{fig:4sls-layer-p} shows the resulting LSHC and
layer-resolved polarization profile in 4 SLs $\MBT$ with the N\'eel vector oriented along the $z$ axis.
The $x$ component is negligible in all four layers,
whereas $P_y$ alternates in sign between adjacent SLs, consistent with the staggered AFM order. 
By contrast, $P_z$ has the same sign in
each surface layer and its adjacent subsurface layer, reverses between the two inner layers, and has opposite signs on the top and bottom surfaces.
Even in the high symmetric configuration with the N\'eel vector oriented along the $z$ axis, 
the layer-resolved polarization of SHC is tilted in $yz$ plane because of the lack of the mirror plane along $xz$.
It is expected that when $\hat{\mathbf m}$ deviates from the $z$ axis, the lower magnetic symmetry allows richer spin polarization from the magnetization-dependent LSHC.

\begin{table*}[t]
\caption{\label{table:fit}
Fitted coefficient vectors (in units of $e/(4\pi)$) with local response axis tilt angles for the total response of a 2 SLs $\MBT$ and the layer-resolved contributions of a 4 SLs $\MBT$ (surface and inner layers).
}
\begin{ruledtabular}
\begin{tabular}{lcccc}
System / Layer & Tilt angle $\theta$ & $\boldsymbol{\sigma}_{\text{SHE0}}$ & $\boldsymbol{\sigma}_{\text{SAHE1}}$ & $\boldsymbol{\sigma}_{\text{SAHE2}}$ \\
\colrule
2 SLs (Total) & $0^\circ$ & $(0, 0.65, -0.12)$ & $(0, 0.36, 0.19)$ & $(0, 0, 0)$ \\
4 SLs surface & $\approx 22^\circ$ & $(0, 0.63, -0.23)$ & $(0, 0.53, 0.30)$ & $(0.18, 0.11, -0.18)$ \\
4 SLs inner & $\approx 9^\circ$ & $(0, 0.12, -0.42)$ & $(0, 0.77, 0.20)$ & $(0.29, 0.23, -0.11)$ \\
\end{tabular}
\end{ruledtabular}
\end{table*}

\begin{figure}
\includegraphics[width=1.0\columnwidth]{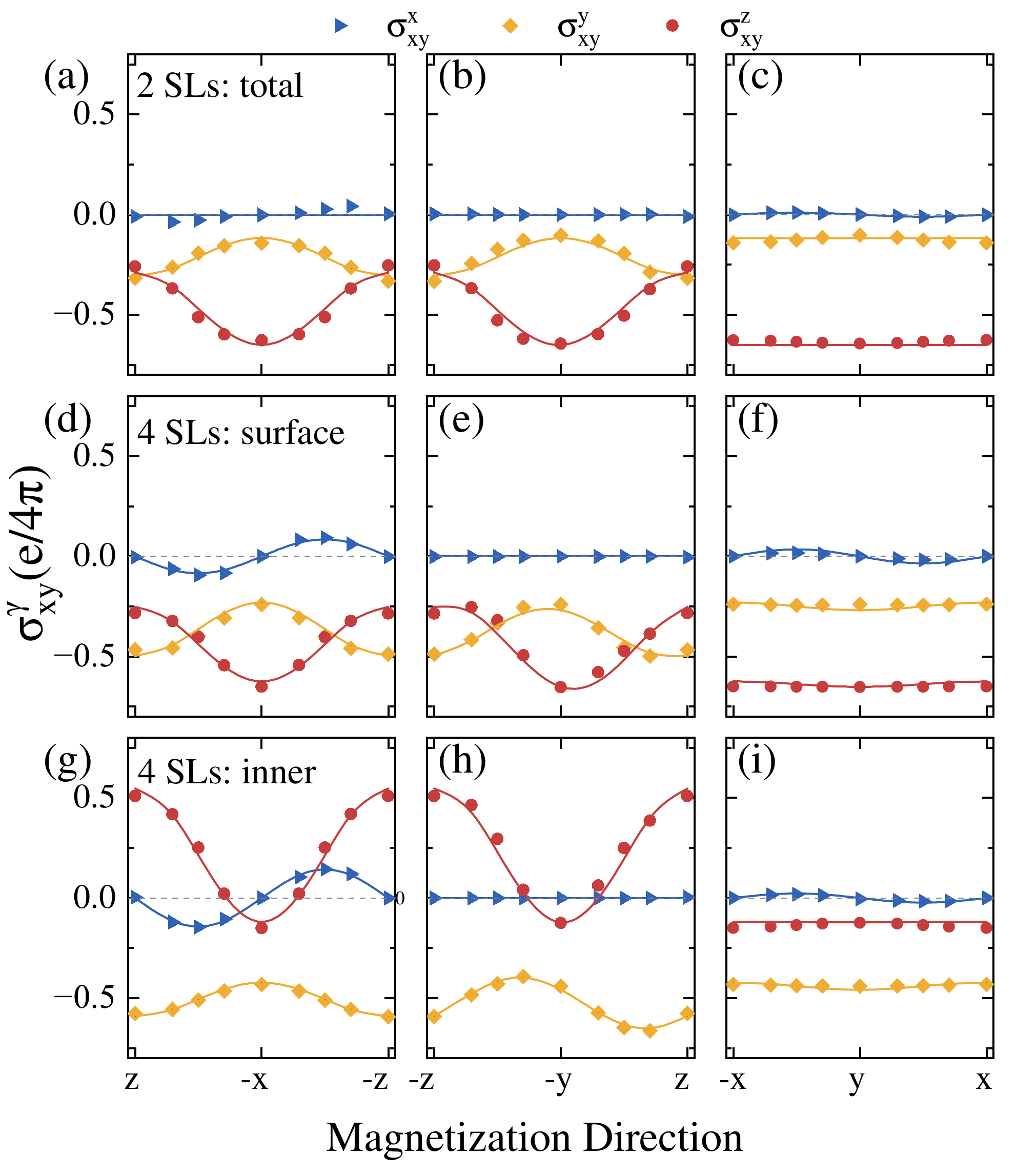}
\caption{
Magnetization dependence of the in-gap SHC $\sigma^{\gamma}_{xy}$ for 2 SLs and 4 SLs $\MBT$. 
(a)-(c) Total SHC of 2 SLs $\MBT$; 
(d)-(f) and (g)-(i) detail the separated SHC contributions from the surface (layers 1 and 4) and inner (layers 2 and 3) layers of 4 SLs $\MBT$, respectively.
Columns correspond to the N\'eel vector rotated within the $xz$, $yz$, and $xy$ planes.
Blue triangles, orange diamonds, and red circles denote $\gamma=x,y,z$, respectively.
Symbols denote the first-principles results, while lines show the fits.
}
\label{fig:m-shc} 
\end{figure}

LSHC inside the topological band gap dependent on the N\'eel vector direction is then investigated in both 2 SLs and 4 SLs $\MBT$.
As shown in Fig.~\ref{fig:m-shc}(a)--(c), in the 2 SLs case, 
the $x$ component of SHC $\sigma_{xy}^{x}$ remains zero for all N\'eel vector directions.
$\sigma_{xy}^{y}$ and $\sigma_{xy}^{z}$ exhibit the same symmetric angular
dependence with the rotation of the N\'eel vector in the $xz$ or $yz$ plane.
Specifically, $\sigma_{xy}^{y}$ varies from approximately
$-0.33\,e/(4\pi)$ for $\mathbf{m}\parallel\pm\hat{\mathbf z}$ to
$-0.10\,e/(4\pi)$ for in-plane orientations, whereas
$\sigma_{xy}^{z}$ changes from approximately
$-0.26\,e/(4\pi)$ for $\mathbf{m}\parallel\pm\hat{\mathbf z}$ to
$-0.64\,e/(4\pi)$ for in-plane orientations.
This indicates that the SHC in the 2 SLs case is almost independent of the in-plane N\'eel vector components,
which is also verified in Fig.~\ref{fig:m-shc}(c).
Considering that the SHC is periodic in the zenith angle and independent of the azimuthal angle, 
we can thus obtain the magnetization-dependent spin anomalous Hall conductivity (SAHC) as:
$(\hat{\mathbf{m}} \cdot \hat{\mathbf{z}})^2 (\boldsymbol\sigma_\text{SAHE1} \times \hat{\mathbf{x}})$.
This term is an even function of $\hat{\mathbf{m}}$.
The corresponding coefficient vector $\boldsymbol\sigma_\text{SAHE1}$ is given in Table~\ref{table:fit}.


For 4 SLs $\MBT$, the LSHC contributions from the surface layers (1 and 4) [Fig.~\ref{fig:m-shc}(d)--(f)] and the inner layers (2 and 3) [Fig.~\ref{fig:m-shc}(g)--(i)] are investigated separately.
In contrast to 2 SLs,  4 SLs $\MBT$ develops a finite
$\sigma_{xy}^{x}$ component in the $xz$ rotation plane.
$\sigma_{xy}^{x}$ is an odd function with respect to the $-x$ direction of $\hat{\mathbf{m}}$ in the $xz$ plane.
In the $yz$ rotation plane in which $\sigma_{xy}^{x}$ is zero, 
the symmetry point of $\sigma_{xy}^{y}$ and $\sigma_{xy}^{z}$ shows a clear offset 
from the principal crystallographic directions.
It is due to the in-equivalence of the $xz$ and $yz$ plane in $\MBT$.
With the N\'eel vector lying in the $xz$ plane, 
the in-plane magnetic moment component is invariant under the mirror symmetry operation $M_{yz}$.
In contrast, the in-plane component of the N\'eel vector lying in the $yz$ plane changes the sign under the mirror symmetry $M_{yz}$. 
Consequently, the symmetry of the total SHC is broken in the $yz$ rotation plane.
Therefore, the offset is described by a local response axis $\mathbf{z}'$
tilted within the $yz$ plane and the corresponding SAHC is given by:
\begin{eqnarray}
\boldsymbol{\sigma}_{xy}^{\text{SAHE}} &=& - (\hat{\mathbf{m}} \cdot \hat{\mathbf{z}'})^2 (\boldsymbol\sigma_\text{SAHE1} \times \hat{\mathbf{x}}) \nonumber \\
& &+ (\hat{\mathbf{m}} \cdot \hat{\mathbf{z}'}) \boldsymbol{\sigma}_{\text{SAHE2}} \odot \hat{\mathbf{m}}
.\label{eq:sahe-4sl}
\end{eqnarray} 
The $\boldsymbol{\sigma}_{\text{SAHE2}}$ term takes a form similar to the SAHE found in certain ferromagnetic materials\cite{PhysRevResearch.3.L042033,PhysRevB.99.220405}.
Fits to Eq.~(\ref{eq:sahe-4sl}) yield tilt
angles of approximately $22^\circ$ for the surface-layer contribution
and $9^\circ$ for the inner-layer contribution
(Table~\ref{table:fit}).
The results of coefficient vectors in Table~\ref{table:fit} show that the inner layers exhibit a larger magnetization-dependent spin Hall response than the surface layers, including 2 SLs cases. 
This also indicates that the topological surface states themselves are not only insensitive to magnetic impurities, 
but the SHC arising from them is also insensitive to the magnetic direction or N\'eel vector direction. 
This is similar to the magnetic insensitive SHC in the magnetic topological material $\text{Mn}\text{Sb}_{2}\text{Te}_{4}$\cite{li2025magnetization}.

\begin{figure}
\includegraphics[width=1\columnwidth]{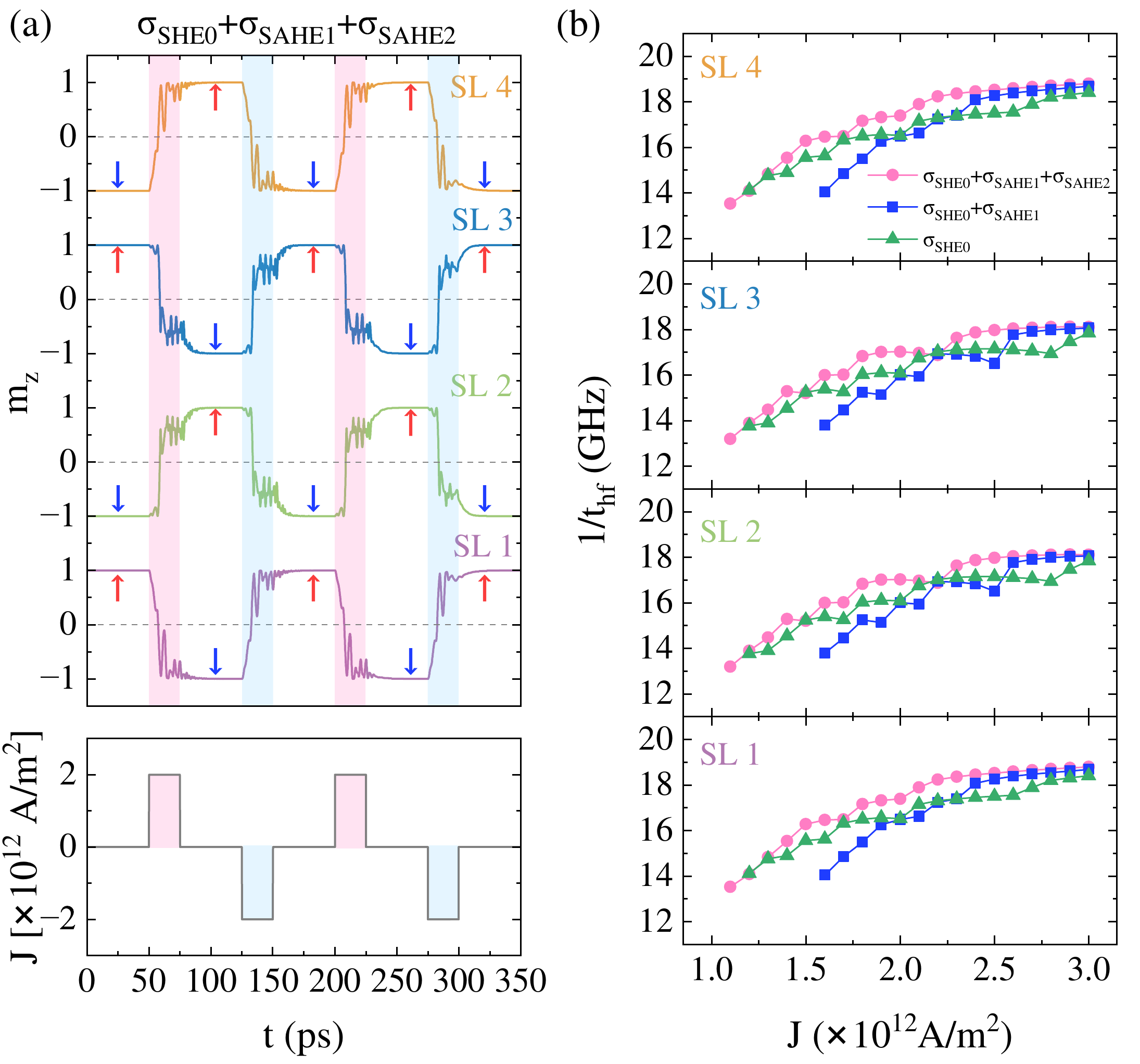}
\caption{(a) Layer-resolved N\'eel vector switching driven by LSOT in 4 SLs $\MBT$.
 The switching is induced by the complete spin response tensor, comprising $\boldsymbol{\sigma}_{\text{SHE0}}$, $\boldsymbol{\sigma}_{\text{SAHE1}}$ and $\boldsymbol{\sigma}_{\text{SAHE2}}$.
The upper panels show the time evolution of the out-of-plane magnetization component $m_{iz}$ ($i=1$--$4$) in each SL under periodic bipolar current pulses with amplitudes of $\pm 2.0\times10^{12}\,\mathrm{A\,m^{-2}}$, whereas the lower panels show the corresponding current density.
(b) the spin-flip frequency $f_{\mathrm{flip}}$ as a function of the spin current density under the situations that $\boldsymbol{\sigma}_{\text{SHE0}}$, $\boldsymbol{\sigma}_{\text{SAHE1}}$ and $\boldsymbol{\sigma}_{\text{SAHE2}}$ is included, only $\boldsymbol{\sigma}_{\text{SHE0}}$, $\boldsymbol{\sigma}_{\text{SAHE1}}$ is included, and only $\boldsymbol{\sigma}_{\text{SHE0}}$ is included, respectively.}
\label{fig:sahe-sot} 
\end{figure}
To examine the dynamical consequences of the layer-resolved spin response, we performed spin dynamics simulations of 4 SLs $\MBT$ using GPU-supported micromagnetic packages MicroMagnetic.jl\cite{wang2024micromagnetic}.
Fig.~\ref{fig:sahe-sot}(a) shows the layer-resolved spin dynamics of 4 SLs $\MBT$ under periodic bipolar spin current pulses. 
the spin current generated by the LSHE for each SL is denoted by $\mathbf{u}_{s}(L)=\frac{J\,S_{xz}}{e^{2}}\boldsymbol\sigma_{xy}^{\mathrm{LSHE}}(L) \mathrm{sgn}(\mathrm{\sigma}_{xy}^{\mathrm{LAHE}}(L))$, $J$ is the Hall current density and $S_{xz}$ is the cross-sectional area perpendicular to the current direction.
A positive current pulse switches the N\'eel vector of each layer along the spin-polarization direction, 
converting the initial $\uparrow\downarrow\uparrow\downarrow$ configuration into the reversed $\downarrow\uparrow\downarrow\uparrow$ state.
Once the pulse is turned off, the interlayer antiferromagnetic exchange coupling drives further relaxation and stabilizes this switched configuration.
A subsequent negative current pulse then flips the layer magnetization into the opposite direction, eventually restoring the initial $\uparrow\downarrow\uparrow\downarrow$ state.
The transient oscillations during the switching process reflect the dynamic competition between current-driven reorientation and interlayer exchange relaxation\cite{gomonay2010spin,cheng2015ultrafast}.
Spin dynamic simulations of 6 SLs $\MBT$ provide similar results [see details in the Supplemental Material].

We then employ the spin-flip frequency as
$f_{\mathrm{flip}}=1/t_{\mathrm{hf}}$, where $t_{\mathrm{hf}}$ is the half spin-flip time to quantify the rate of magnetic domain switching.
As shown in Fig.~\ref{fig:sahe-sot}(b), the $f_{\mathrm{flip}}$ increases as the driving current increases, 
indicating that the magnetic domains switch faster under a larger SOT. 
However, it saturates at about $3\times10^{12}\,\mathrm{A\,m^{-2}}$, 
because this frequency is largely determined by the relaxation of the domains after the SOT is removed.
The complete $\boldsymbol{\sigma}_{\text{SHE0}}+\boldsymbol{\sigma}_{\text{SAHE1}}+\boldsymbol{\sigma}_{\text{SAHE2}}$ spin response provides the highest $f_{\mathrm{flip}}$ and the lowest critical current density for domain flipping.
Further investigation on the 
$\boldsymbol{\sigma}_{\text{SHE0}}+\boldsymbol{\sigma}_{\text{SAHE1}}$ LSOT exhibits a higher critical current density.
It is due to the pronounced oscillations developing in the layer-resolved N\'eel vector during the current pulse, 
whereas the trajectory is much smoother when $\boldsymbol{\sigma}_{\text{SAHE1}}$ is omitted.
The detailed studies are shown in the Supplemental Material.
Overall, the switching efficiency is governed by layer dependence of the spin response, with the complete response enabling the fastest reversal.

It is worth noting that the simulations above utilized the spin current generated by LSHE, rather than the charge current. 
Given that the charge conductivity approaches zero within the topological band gap, 
the magnitude of the actual charge current employed would be one to two orders of magnitude smaller than that of the spin Hall current. 
Although precise experimental observation of the quantum anomalous Hall effect in two-dimensional magnetic topological materials remains challenging, 
the large spin Hall angle can persist even in the presence of neutral defects, 
owing to the robustness of topological edge states\cite{li2025magnetization}. 
In other words, research on magnetic-related LSHE not only provides new insights for achieving low-power, high-efficiency spintronic control using two-dimensional magnetic topological materials, 
but also serves as an effective way to identify the topological edge states of magnetic topological materials.

\begin{acknowledgments}
\emph{Acknowledgments ---}
This work was financially supported by the National Natural Science
Foundation of China (12274309).

\end{acknowledgments}

\bibliographystyle{apsrev4-2}

\end{document}